\documentclass[11pt]{article}

\usepackage[preprint]{acl}

\usepackage{times}
\usepackage{latexsym}

\usepackage[T1]{fontenc}

\usepackage[utf8]{inputenc}

\usepackage{inconsolata}

\usepackage{microtype}
\usepackage{graphicx}
\usepackage{subcaption}
\usepackage{booktabs} 
 \usepackage{multirow} 
\usepackage{tabularx}
\usepackage{xcolor}
\usepackage{makecell}
\usepackage{bbding}
\usepackage{algorithmic}
\usepackage{algorithm}
\usepackage{tcolorbox}
\usepackage{fvextra}
\usepackage{amsmath}
\usepackage{amssymb}
\usepackage{mathtools}
\usepackage{amsthm}

\usepackage[capitalize,noabbrev]{cleveref}

\theoremstyle{plain}

\theoremstyle{definition}

\theoremstyle{remark}

\title{Symbolic-Neural Soft-Logic Reasoning: Towards Robust and Verifiable Thinking Chains via Cooperative Evolution}

\author{Rui Wang$^{1*}$, Zeming Wei$^{1}$, Yihao Zhang$^{1}$, Xiaokun Luan$^{1}$\\
  $^{1}$School of Mathematical Sciences, Peking University / Beijing, China\\
  \texttt{2300010607@stu.pku.edu.cn}
  }

\begin{document}
\maketitle




\begin{abstract}
Large Language Models (LLMs) have demonstrated impressive progress in complex reasoning tasks, largely driven by the Chain-of-Thought (CoT) paradigm, which decomposes difficult problems into intermediate steps. However, CoT reasoning remains fundamentally constrained by the probabilistic nature of neural generation, leading to unfaithful reasoning chains that undermine reliability. Neuro-symbolic approaches attempt to address these issues by combining LLMs with symbolic solvers, yet they face persistent challenges, including hallucinated translations, the mismatch between natural language and formal logic, and the limited enhancement of the LLM’s intrinsic reasoning ability. To overcome these limitations, we propose Symbolic-Neural Soft-Logic Reasoning (SSR), a unified framework that integrates LLMs with symbolic reasoning and improves robustness by relaxing strict logical determinism while preserving verifiability. Our approach improves reasoning performance, automatically generates verifiable and human-like logical thinking chains for training and fine-tuning, and facilitates cross-disciplinary applications such as AI for mathematics. Experiments across multiple models and benchmarks demonstrate that SSR consistently outperforms existing reasoning frameworks, highlighting its effectiveness in enhancing both the robustness and interpretability of LLM reasoning.

\end{abstract}
\section{Introduction}

Large Language Models (LLMs) have achieved remarkable progress on complex reasoning tasks, exemplified by recent systems such as DeepSeek-R1~\cite{guo2025deepseek} and GPT-4o~\cite{hurst2024gpt}. Reasoning in LLMs broadly refers to the ability to maintain logically consistent internal states, integrate world knowledge, and perform multi-step deductions over long contexts~\cite{plaat2024reasoning}. However, vanilla autoregressive generation suffers from a fundamental \emph{one-step bottleneck}: for complex multi-step reasoning problems, LLMs struggle to maintain a correct line of reasoning across a long sequence of dependent token predictions, often resulting in premature abandonment, random guessing, or getting trapped in local loops.~\cite{arbuzov2025beyond}.

Chain-of-Thought (CoT) prompting~\cite{wei2022chain} has emerged as a powerful paradigm to mitigate this issue by decomposing complex reasoning into explicit intermediate steps, effectively expanding the model’s computational workspace and substantially improving performance, even with minimal prompting~\cite{kojima2022large}. Building upon CoT, structured reasoning frameworks such as Tree of Thoughts~\cite{yao2023tree} and Program of Thoughts~\cite{chen2022program} further enhance reasoning via branching exploration or external execution. Beyond accuracy gains, explicit reasoning chains also provide interpretability and valuable supervision signals for training and fine-tuning.

Despite these advantages, CoT-based reasoning remains fragile. Recent work has shown that reasoning chains may be unfaithful, serving as post-hoc rationalizations rather than reflecting the true causal process~\cite{turpin2023language}, and are susceptible to confirmation bias~\cite{wan2025unveiling}, where models tend to reinforce their initial priors during the reasoning process. More fundamentally, LLM reasoning inherits hallucination issues rooted in probabilistic generation, including fabrication, incorrect inference, and misuse of rules~\cite{sun2025detection}. These issues compromise both the reliability and educational value of LLM-based reasoning systems, even when final answers are correct.
\begin{figure*}
    \centering
    \includegraphics[width=1\linewidth]{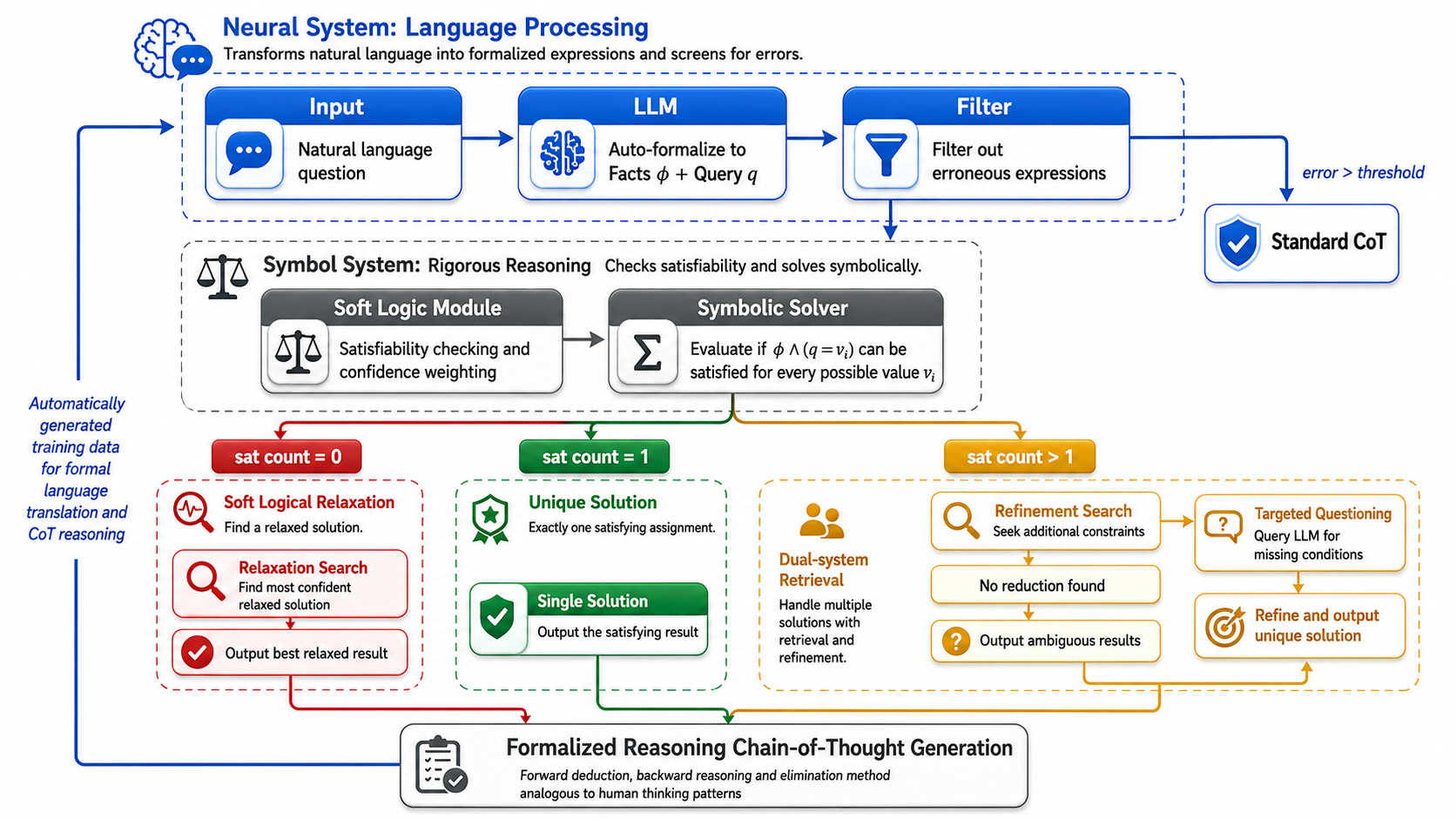}
    \caption{Overall Process and Application Framework Diagram of the SSR Method}
    \label{fig:pipeline}
\end{figure*}

To address these limitations, neuro-symbolic frameworks combine LLMs with symbolic solvers, leveraging the flexible natural language processing capabilities of neural systems and the rigorous reasoning capabilities of symbolic systems~\cite{hitzler2022neuro,hitzler2022neuro2}. In this paradigm, LLMs translate natural language into symbolic constraints, which are then solved deterministically. While effective in improving correctness, this approach introduces new challenges. Auto-formalization itself is prone to hallucinations, omissions, and structural errors~\cite{weng2025autoformalization}. Moreover, the inherent ambiguity and context dependence of natural language conflict with the strict semantics required by formal logic, often leading to brittle or unsatisfiable formulations. Finally, delegating the core reasoning process entirely to symbolic solvers raises concerns about whether such systems genuinely enhance the LLM’s intrinsic reasoning ability, rather than merely bypassing it~\cite{pan2023logic,feng2025vericot}.

In this work, we argue that these limitations stem from treating symbolic reasoning as strictly deterministic, leaving no tolerance for uncertainty or imperfect formalization. We propose \textbf{Symbolic-Neural Soft-Logic Reasoning (SSR)}, a neuro-symbolic framework that integrates \emph{soft logic} constraints to enhance robustness under noisy, incomplete, or uncertain premises. By allowing constraints to be satisfied approximately rather than absolutely, SSR mitigates the cascading failures caused by local errors. Moreover, SSR introduces deeper interaction between neural and symbolic components, enabling the automatic generation of verifiable, human-like reasoning chains via forward and backward deduction. Our framework aims to (i) improve reasoning accuracy of existing LLMs, (ii) automatically generate faithful and interpretable logical reasoning chains for training and evaluation, and (iii) support practical reasoning-intensive applications such as planning and problem solving in finite discrete mathematics. Experimental results across multiple models and benchmarks demonstrate the effectiveness of SSR on both reasoning accuracy and reasoning-chain quality. Figure~\ref{fig:pipeline} illustrates the overall SSR pipeline.

\section{Methodology}
\label{sec:methodology}

SSR aims to combine the semantic flexibility of LLMs with the consistency guarantees of symbolic solvers. Given a natural language problem, an LLM first abstracts it into a restricted logical program, and an SMT solver then performs verifiable inference over the translated constraints. Since LLMs may introduce errors when translating formal languages, or the inherent properties of the questions prevent a unique answer from being obtained, SSR further introduces soft logical relaxation and identifies chain‑breaking points in dual‑system interactive retrieval reasoning. Moreover, SSR can generate formalized reasoning chains to be translated into concise natural‑language reasoning chains or used as training data.

\subsection{Hard Logical Reasoning}

We first consider an idealized setting where the LLM translates the natural language input into formal logic without ambiguity or error. This setting isolates the core inference problem: once the symbolic program is correct, reasoning can be reduced to deterministic SMT solving. Detailed abstraction rules are provided in Appendix~\ref{app: symbol}.

Given a problem, the LLM produces a set of symbolic expressions consisting of facts and queries. Facts encode known constraints, while queries specify the target predicate functions to be inferred. Predicate functions represent object properties and relations, and are categorized into Numeric Sort or Boolean Sort according to whether their domains are finite enumerable values or truth values, such as $\text{Age}(\text{Bob})=16$ and $\text{Happy}(\text{Anne})$. All expressions follow a restricted SMT-compatible syntax to ensure consistency and verifiability.

Let the fact set be:
\begin{equation}
\Phi = \bigwedge_{i=1}^{n} \varphi_i .
\end{equation}
Let $Q=\{q_1,\dots,q_m\}$ be the query set. Reasoning is performed by checking satisfiability under different query conditions, as detailed in Appendix~\ref{app: procedure}. For a Boolean query $q_i$, SSR checks $\Phi \wedge q_i$ and $\Phi \wedge \lnot q_i$. If only one case is satisfiable, the corresponding truth value is returned; if both are satisfiable, the result is Unknown.

For a numeric query with free predicate variable $a_i$, where $a_i \in \{a_{i1},a_{i2},\ldots,a_{it_i}\}$, SSR checks each candidate assignment $\Phi \wedge (a_i=a_{ij})$. The solution set is:
\begin{equation}
\{v \in \mathcal{D}_i \mid \Phi \wedge (a_i=v)\ \text{is \textbf{sat}}\}.
\end{equation}
Arithmetic reasoning is handled by translating arithmetic operators into symbolic expressions.

Under the idealized assumption, $\Phi$ is satisfiable and at least one query value is satisfiable. Thus, hard logical reasoning provides a rigorous deterministic foundation: inference is delegated to the solver, which enforces global consistency and avoids hallucinated reasoning. Its limitation is equally clear: if the LLM-generated logic is noisy, incomplete, or over-constrained, hard logic becomes brittle.

\subsection{Soft Logical Relaxation}

Hard logic fails when translated constraints are not perfectly reliable. In LLM-assisted reasoning, errors may arise from omitted premises, imprecise predicates, missing negations, or implicit commonsense assumptions. Some real-world conditions are also approximate or context-dependent. SSR therefore relaxes hard constraints into confidence-weighted soft logic, allowing symbolic reasoning to proceed under noise and uncertainty.

Soft logic extends Boolean logic by associating constraints with real-valued confidence scores rather than enforcing all rules as strictly true~\cite{pryor2022neupsl}. This makes reasoning more robust when symbolic rules are imperfect, contradictory, or probabilistic, and has been widely used to connect logical reasoning with statistical learning~\cite{dubois1998soft,kimmig2012short}. SSR adopts this relaxation to tolerate translation errors, missing premises, and uncertain real-world knowledge~\cite{kimmig2012short}.

Before solving, SSR applies a syntax-level filter to the LLM-generated expressions. Clearly malformed expressions, such as raw natural-language fragments, invalid predicate usage, or garbled characters, are replaced by $\mathtt{BoolVal(True)}$, a non-informative placeholder that prevents parser failure. If $\Phi$ contains more than one such placeholder, or if the query $q_i$ is replaced by $\mathtt{BoolVal(True)}$, the translation is treated as severely corrupted and SSR falls back to a conventional CoT pipeline.

For the remaining facts, each $\varphi\in\Phi$ receives a confidence weight $w(\varphi)>0$. We use either a uniform prior or an entropy-based certainty score:
\begin{equation}
w(\varphi)=\exp(-H(\varphi)),
\end{equation}
where $H(\varphi)$ is the length-normalized entropy of the LLM output associated with $\varphi$, computed from token-level logit probabilities. This is because it is negatively correlated with internal uncertainty and unaffected by sentence length. The weight measures how much the model trusts a constraint when the fact set is in tension.

Given weighted facts $\{\varphi_1,\dots,\varphi_n\}$ and query $q_i$, SSR enumerates candidate values $q_{ij}$. For closed logical queries, the candidates are $\{\text{True},\text{False}\}$; for variable queries, they are values in the domain. SSR first checks whether $\Phi$ is satisfiable. If $\Phi$ is $\texttt{unsat}$, it searches for a satisfiable subset with the largest total confidence weight.

After relaxation, if exactly one candidate is $\texttt{sat}$, SSR returns it as in hard logic. If no candidate is $\texttt{sat}$, then for each candidate $(q_i=q_{ij})$, SSR retrieves a subset $\Phi'_j\subseteq\Phi$ such that $\Phi'_j \wedge (q_i=q_{ij})$ is $\texttt{sat}$ and the total confidence weight of $\Phi'_j$ is maximized. SSR selects the candidate associated with the maximum-weight satisfiable subset. If both $\text{True}$ and $\text{False}$ are selected for a Boolean query, the answer is set to \texttt{Unknown}.

Thus, soft relaxation keeps SMT satisfiability as the validity criterion while replacing all-or-nothing failure with confidence-weighted constraint selection in real‑world scenarios where formal logic may not be perfect.

\subsection{Dual-System Targeted Retrieval under Soft Logical Ambiguity}

Soft relaxation resolves inconsistency, but it may still leave multiple candidates satisfiable. This may stem from translation omissions and errors, or the absence of decisive conditions caused by unrecognized implicit information in questions or commonsense knowledge. SSR handles this ambiguity through targeted retrieval: the solver first identifies which predicate assignment would make the answer unique, and the LLM then verifies whether that assignment is supported by the problem statement or commonsense knowledge. 

When more than one candidate is $\texttt{sat}$, SSR enumerates possible predicate-function values and searches for an assignment $a=\text{value}$ such that fixing it makes exactly one candidate satisfiable:
\begin{equation}
\Phi \wedge (a=\text{value}) \wedge (q_i=q_{ij}).
\end{equation}
If no such assignment exists, SSR retains the original results. If such an assignment exists, SSR further searches for another assignment $a'=\text{value}'$ that can derive it. Specifically, SSR checks whether $\Phi \wedge (a'=\text{value}') \wedge (a=\text{value})$ is $\texttt{sat}$ while $\Phi \wedge (a'=\text{value}') \wedge \neg(a=\text{value})$ is $\texttt{unsat}$. If so, $a=\text{value}$ is replaced by $a'=\text{value}'$, and the search continues iteratively until no further retrievable assignment exists. Essentially, this recursion conducts backward search for required information to identify conditions that can connect coherently with forward reasoning, thereby completing the reasoning chain.

The final retrieved assignment is then passed to the LLM. If the LLM confirms that it can be inferred from the problem statement or commonsense knowledge, SSR selects the unique solution corresponding to that assignment. Otherwise, SSR keeps the original set of satisfiable solutions. In this way, the solver localizes the ambiguity, while the LLM only verifies a specific retrieved premise instead of directly choosing the answer.

\subsection{Human-like Chain-of-Thought Auto-Generation}
\begin{figure}
    \centering
    \includegraphics[width=1\linewidth]{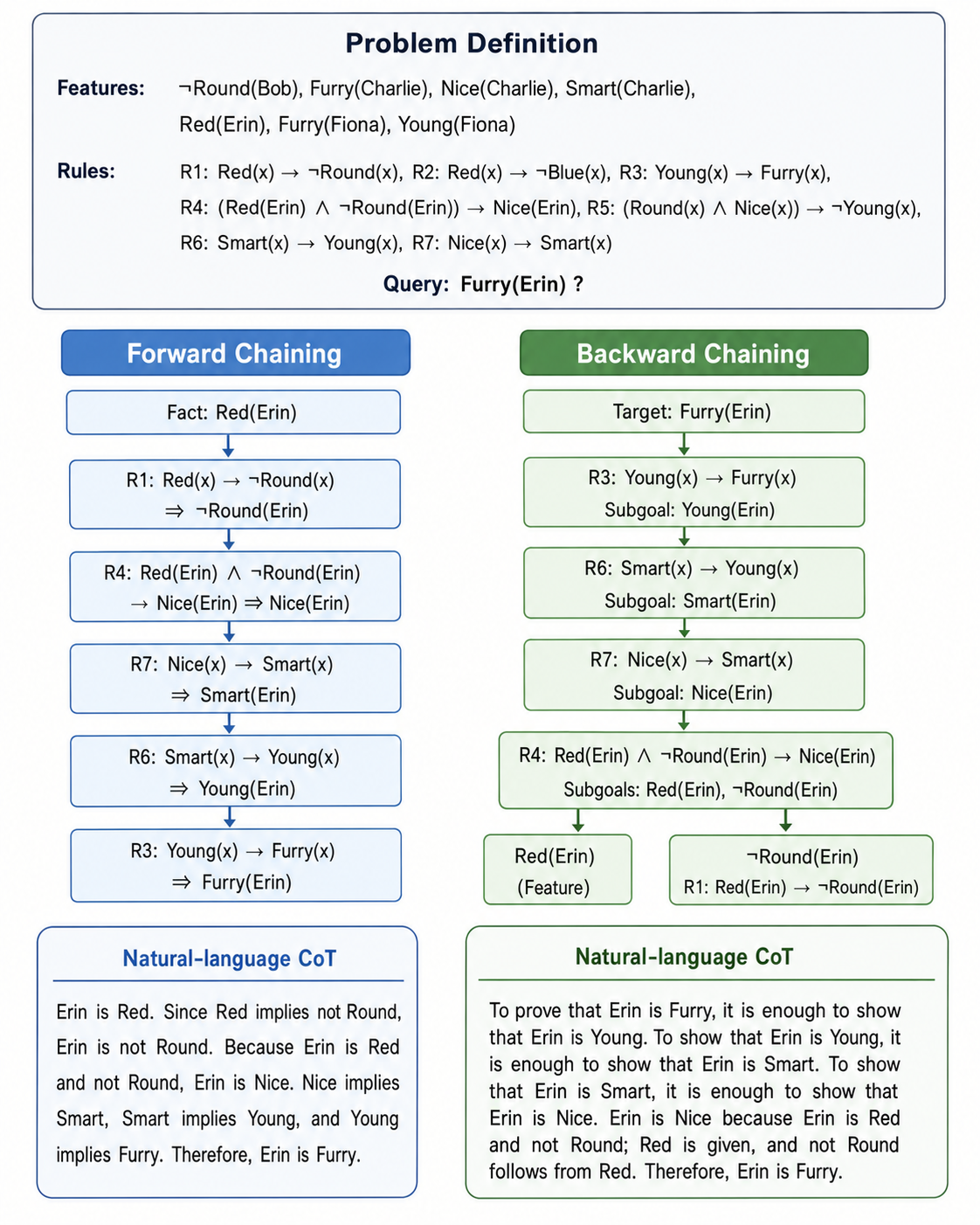}
    \caption{Example reasoning chains generated via forward and backward deduction in the SSR framework.}
    \label{fig:chain}
\end{figure}
SMT solving yields verifiable answers but not human-readable reasoning traces. To improve interpretability and practicality, SSR constructs symbolic reasoning chains from the verified logical program. The generated explanation is therefore grounded in symbolic inference rather than randomly generated by the LLM.

SSR first normalizes symbolic inputs into properties and rules. A property specifies that a predicate assignment is given or excluded, such as $\text{Predicate}(\text{Object})$ or $\neg\text{Predicate}(\text{Object})$. A rule encodes an implication between properties, such as $P \rightarrow Q$. Quantified statements are expanded into conjunctions or disjunctions of simple rules. If the query itself is a rule, its premise $P$ is temporarily added to the fact set and its conclusion $Q$ becomes the new target.

Backward reasoning starts from the query target and retrieves rules whose conclusions match it. The premises of these rules become sub-goals, and the process repeats until the target is supported by given properties or no further rule applies. If the query is determined by excluding alternatives, this becomes backward reasoning with elimination, which subsumes ``proof by contradiction''.

Forward reasoning starts from known properties and iteratively applies rules whose premises are satisfied. Newly derived properties are added to the knowledge set and may trigger further rules. The process stops when the query predicate is determined, either directly or by elimination, or when no new property can be derived.

The resulting forward and backward traces form explicit symbolic chains. These chains can be stored as upright and inverted n-ary trees, respectively, translated by an LLM into natural-language CoT, grouped by reasoning pattern, and can be automatically verified, with the shortest paths selected separately as the standard answers for different methods. Besides interpretability, this provides scalable and verifiable reasoning traces, reducing dependence on costly human annotation~\cite{ahn2024large} and addressing a limitation of prior neuro-symbolic systems, where symbolic reasoning improved accuracy but did not help the model improve its intrinsic reasoning capability~\cite{pan2023logic}.

Overall, SSR follows a progressive design: neural system processes diverse languages, symbolic system provides deterministic inference, soft logic addresses the fragility of formal deduction, targeted retrieval resolves ambiguity and clue breakage, and symbolic CoT generation exposes the reasoning process in a verifiable human-readable form.
\section{Experiments}

\subsection{Experimental Setup}

\noindent \textbf{Models.} We evaluate our method using a diverse set of LLMs. Specifically, we include two strong open-source instruction-tuned models, \textbf{Qwen2.5-7B-Instruct}~\cite{yang2025qwen2} and \textbf{DeepSeek-R1}~\cite{guo2025deepseek}, as well as the proprietary model \textbf{GPT-4o}~\cite{hurst2024gpt} and \textbf{GPT-5.4}~\cite{singh2025openai}. This selection allows us to assess the effectiveness of our framework across models with different architectures and scales.

\noindent \textbf{Datasets.} We conduct experiments on multiple benchmark datasets covering a broad spectrum of logical reasoning tasks. \textbf{PrOntoQA}~\cite{saparov2022language} is used to evaluate deductive reasoning capabilities under well-structured logical settings. While \textbf{ProofWriter}~\cite{tafjord2021proofwriter} extends deductive reasoning by introducing an \emph{unknown} option, testing whether LLMs can correctly assess the sufficiency of given conditions. \textbf{FOLIO}~\cite{han2024folio} contains more complex first-order logic reasoning problems, where long and syntactically complex sentences as well as implicit commonsense assumptions pose significant challenges to LLM-based reasoning. Finally, \textbf{LogicalDeduction} is a logical reasoning task from the BigBench benchmark~\cite{ghazal2013bigbench} used to evaluate constraint satisfaction problems in discrete combinatorial scenarios.

\noindent \textbf{Baselines.} We compare our proposed \textbf{SSR} method against several representative baselines. These include \textbf{Direct} answering, where the model generates an answer without explicit reasoning; standard Chain-of-Thought (\textbf{CoT})~\cite{wei2022chain} prompting; \textbf{Logic-LM}~\cite{pan2023logic}, which translates natural language into formal logic and delegates reasoning to a symbolic solver; and \textbf{VeriCoT}~\cite{feng2025vericot}, which first generates a natural-language CoT and then verifies it step by step using a symbolic solver. These baselines cover classical prompt-engineering approaches and two representative neuro-symbolic reasoning frameworks. In addition, we introduce a self-designed \textbf{Formal CoT} baseline, where the LLM is prompted to generate a purely formal logical reasoning chain. We do not include other popular paradigms such as implicit reasoning or supervised fine-tuning, as these approaches are not readily applicable to closed-source models.

\noindent \textbf{Evaluation Protocol.} We set the maximum token generation length to 1000 and use the \emph{Z3-solver}~\cite{de2008z3} as the symbolic reasoning backend. For each dataset, we sample 200 problems for evaluation. To measure improvements in LLM reasoning performance, we compare the final-answer accuracy and the total inference time across all methods. For evaluating generated reasoning chains, we require not only that the final answer be correct, but also that each step in the reasoning chain be strictly justified by the problem statement, valid commonsense assumptions, or logically derived conclusions. Since the Direct and Logic-LM baselines do not produce explicit reasoning chains, we only compare reasoning-chain quality against \textbf{Formal CoT} and \textbf{VeriCoT}. 

\subsection{Main Results}
\begin{table*}
\caption{Overall comparison of SSR vs. Baselines on improving reasoning accuracy of different models across datasets. Here, FCoT refers to Formal CoT, L-LM refers to Logic-LM, and V-CoT refers to VeriCoT. All data in the table represent the mean ± standard deviation over five repeated experiments.}
\label{tab: accuracy}
\small
\begin{tabularx}{\linewidth}{|l|l|*{6}{>{\centering\arraybackslash}X|}}
\toprule
Model & Dataset & Direct & CoT & FCoT & L-LM & V-CoT & SSR\\
\midrule
\multirow{4}{*}{Qwen-2.5-7B} 
& PrOntoQA 
& 67.5$_{\pm 0.00}$
& 86.1$_{\pm 0.37}$
& 84.9$_{\pm 0.20}$
& 91.7$_{\pm 0.24}$
& 88.1$_{\pm 0.20}$
& $\textbf{97.4}_{\pm 0.20}$ \\

& ProofWriter 
& 37.0$_{\pm 0.00}$
& 58.1$_{\pm 0.37}$
& 50.9$_{\pm 0.20}$
& 77.9$_{\pm 0.20}$
& 73.4$_{\pm 0.49}$
& $\textbf{86.0}_{\pm 0.32}$ \\

& FOLIO 
& 55.5$_{\pm 0.00}$
& 64.5$_{\pm 0.45}$
& 56.1$_{\pm 0.20}$
& 62.2$_{\pm 0.24}$
& $\textbf{66.0}_{\pm 0.32}$
& 64.6$_{\pm 0.20}$ \\

& LogicalDeduction  
& 46.0$_{\pm 0.00}$
& 63.2$_{\pm 0.4}$
& 41.8$_{\pm 0.24}$
& 72.0$_{\pm 0.32}$
& 66.5$_{\pm 0.40}$
& $\textbf{75.0}_{\pm 0.32}$ \\
\midrule

\multirow{4}{*}{DeepSeek-R1} 
& PrOntoQA 
& 86.5$_{\pm 0.00}$
& 96.0$_{\pm 0.00}$
& 91.7$_{\pm 0.24}$
& 92.0$_{\pm 0.32}$
& 95.4$_{\pm 0.20}$
& $\textbf{100.0}_{\pm 0.00}$ \\

& ProofWriter 
& 56.0$_{\pm 0.00}$
& 72.5$_{\pm 0.32}$
& 73.5$_{\pm 0.77}$
& 89.0$_{\pm 0.00}$
& 80.1$_{\pm 0.20}$
& $\textbf{98.9}_{\pm 0.20}$ \\

& FOLIO 
& 58.5$_{\pm 0.00}$
& 68.0$_{\pm 0.37}$
& 56.2$_{\pm 0.24}$
& 69.5$_{\pm 0.00}$
& 64.4$_{\pm 0.37}$
& $\textbf{71.5}_{\pm 0.32}$ \\

& LogicalDeduction  
& 74.5$_{\pm 0.00}$
& 84.0$_{\pm 0.20}$
& 83.5$_{\pm 0.49}$
& 88.7$_{\pm 0.40}$
& 85.6$_{\pm 0.20}$
& $\textbf{90.2}_{\pm 0.24}$ \\
\midrule

\multirow{4}{*}{GPT-4o} 
& PrOntoQA 
& 83.5$_{\pm 0.00}$
& 94.5$_{\pm 0.00}$
& 86.0$_{\pm 0.00}$
& 88.5$_{\pm 0.00}$
& 96.5$_{\pm 0.00}$
& $\textbf{100.0}_{\pm 0.00}$ \\

& ProofWriter 
& 52.5$_{\pm 0.00}$
& 74.0$_{\pm 0.37}$
& 68.2$_{\pm 0.24}$
& 90.0$_{\pm 0.00}$
& 80.6$_{\pm 0.20}$
& $\textbf{98.0}_{\pm 0.00}$ \\

& FOLIO 
& 64.5$_{\pm 0.00}$
& 70.5$_{\pm 0.32}$
& 57.7$_{\pm 0.24}$
& $\textbf{76.1}_{\pm 0.20}$
& 66.2$_{\pm 0.24}$
& 75.2$_{\pm 0.24}$ \\

& LogicalDeduction  
& 63.5$_{\pm 0.00}$
& 72.0$_{\pm 0.00}$
& 64.4$_{\pm 0.49}$
& 86.0$_{\pm 0.00}$
& 72.6$_{\pm 0.24}$
& $\textbf{88.4}_{\pm 0.24}$ \\
\midrule

\multirow{4}{*}{GPT-5.4} 
& PrOntoQA 
& 97.0$_{\pm 0.00}$
& 98.5$_{\pm 0.00}$
& 90.0$_{\pm 0.32}$
& $\textbf{100.0}_{\pm 0.00}$
& 97.4$_{\pm 0.24}$
& $\textbf{100.0}_{\pm 0.00}$ \\

& ProofWriter 
& 79.5$_{\pm 0.00}$
& 87.1$_{\pm 0.20}$
& 78.4$_{\pm 0.20}$
& 95.0$_{\pm 0.40}$
& 89.2$_{\pm 0.24}$
& $\textbf{98.6}_{\pm 0.58}$ \\

& FOLIO 
& 69.5$_{\pm 0.00}$
& 74.5$_{\pm 0.32}$
& 64.8$_{\pm 0.40}$
& 76.4$_{\pm 0.49}$
& 72.3$_{\pm 0.24}$
& $\textbf{78.1}_{\pm 0.49}$ \\

& LogicalDeduction  
& 71.0$_{\pm 0.00}$
& 76.6$_{\pm 0.20}$
& 69.8$_{\pm 0.24}$
& 88.7$_{\pm 0.40}$
& 79.9$_{\pm 0.37}$
& $\textbf{90.0}_{\pm 0.32}$ \\
\midrule

\multicolumn{2}{|c|}{Total Runtime (min)} & 35.7 & 265.6 & 288.6 & 362.0 & 451.5 & 435.1\\
\bottomrule
\end{tabularx}
\end{table*}
\noindent \textbf{Overall Superiority of SSR in Enhancing Reasoning Accuracy.} As shown in Table~\ref{tab: accuracy}, SSR is the only method that achieves consistent improvements over the standard CoT across all datasets and models. Notably, SSR achieves strong gains even when compared to advanced neuro-symbolic baselines such as Logic-LM and VeriCoT, indicating that its advantages cannot be attributed solely to verification or symbolic execution. These results suggest that the soft-logic integration in SSR enables more robust and flexible multi-step reasoning, effectively mitigating the stochastic and error-accumulation issues inherent in auto-formalization. Overall, the consistent dominance of SSR across models and reasoning tasks highlights its ability to enhance both reasoning accuracy and cross-dataset generalization. For detailed case analyses and discussions of the underlying causes of failures, please refer to Appendix~\ref{app: case}.

\noindent \textbf{Analysis of the Baselines.} As shown in the table, CoT reasoning significantly improves accuracy compared to directly producing final answers. In contrast, Formal CoT often underperforms standard CoT, suggesting that LLMs are more adept at reasoning in natural language than in symbolic logic since the inherently probabilistic nature of LLM generation remains unchanged. Logic-LM achieves notable overall improvements over CoT, while VeriCoT generally benefits from verification feedback to enhance final reasoning accuracy; however, VeriCoT cannot prevent cases where responses with incorrect reasoning but correct answers are modified into outputs with both incorrect reasoning and incorrect final answers.

\noindent \textbf{Generating Correct and Verifiable Thinking Chains.} Table~\ref{tab: chain} demonstrates that SSR is markedly more effective than existing baselines at generating fully correct and verifiable reasoning thought chains across different models and datasets. While VeriCoT rely on post-hoc verification of generated outputs, SSR directly integrates retrieval and verification into the reasoning loop, substantially increasing the likelihood that the generated thought chain is consistent and logically complete. Moreover, it is worth noting that the forward deduction approach is relatively more stable, which may be because retrieving relevant premises based on the goal during backward deduction is more prone to errors. Overall, these results confirm that deep neural–symbolic interaction in SSR enables LLMs to construct verifiable, end-to-end correct reasoning chains rather than merely producing plausible but invalid explanations. This can automatically provide multi-path, verifiable chains of thought for LLM inference training..
\begin{table}
\vspace{-0.4cm}
\caption{Overall comparative experiment on the success rate of generating fully correct reasoning chains between SSR and baselines. We adopt abbreviations for the names of models and datasets, and ``SSR-F'', ``SSR-B'' refer to the variants of the SSR method using forward deduction and backward deduction algorithms, respectively. All data in the table represent the average accuracy over five repeated experiments.}
\label{tab: chain}
\small
\begin{tabularx}{\linewidth}{l|l|*{2}c|*{2}c}
\toprule
Model & Dataset & FCoT & V-CoT & SSR-F & SSR-B\\
\midrule
\multirow{4}{*}{Qwen} 
& PQ & 21.9 & 51.7 & \textbf{89.7} & 88.5\\
& PW & 14.5 & 32.3 & \textbf{63.5} & 62.0\\
& FO & 10.8 & 17.9 & \textbf{34.6} & 31.1\\
& LD & 22.6 & 34.6 & \textbf{55.4} & 50.8\\
\midrule
\multirow{4}{*}{DS} 
& PQ & 33.8 & 67.4 & \textbf{94.5} & \textbf{94.5}\\
& PW & 18.1 & 38.3 & \textbf{73.1} & 72.4\\
& FO & 13.5 & 22.0 & \textbf{39.4} & 35.5\\
& LD & 31.7 & 44.3 & \textbf{69.7} & 63.9\\
\midrule
\multirow{4}{*}{GPT-4o} 
& PQ & 34.6 & 66.7 & \textbf{94.0} & 93.9\\
& PW & 20.7 & 39.7 & \textbf{71.9} & 70.2\\
& FO & 15.4 & 23.0 & \textbf{41.5} & 36.1\\
& LD & 30.1 & 42.7 & \textbf{66.0} & 61.2\\
\midrule
\multirow{4}{*}{GPT-5.4} 
& PQ & 36.6 & 68.3 & \textbf{94.9} & 94.2\\
& PW & 22.9 & 39.5 & \textbf{71.4} & 70.4\\
& FO & 15.2 & 23.4 & \textbf{42.2} & 38.0\\
& LD & 32.5 & 44.3 & \textbf{67.0} & 63.6\\
\midrule
\multicolumn{2}{c|}{Total Time (min)} & 389.6 & 630.4 & 708.5 & 595.3\\
\bottomrule
\end{tabularx}
\vspace{-0.4cm}
\end{table}

\subsection{Ablation Study and Comparative Study}
\noindent \textbf{Attribution of execution-branch accuracy.} As shown in Table~\ref{tab: chi}, the distributions of correct and incorrect answers between the symbolic execution branch and the CoT fallback branch in SSR exhibit statistically significant differences. In particular, samples successfully handled by symbolic execution tend to be far more reliable than those diverted to CoT due to logical translation failures, highlighting the effectiveness of formal verification when symbolic parsing succeeds. These results demonstrate that SSR’s gains primarily stem from its ability to maximize successful symbolic execution, while the CoT branch serves as a necessary but less reliable backup. Moreover, it is worth noting that, within the SSR framework, samples processed via the CoT branch exhibit a lower accuracy than the overall accuracy of the standalone CoT method. This suggests that if an LLM finds it difficult to translate a problem into logical form, the problem is also statistically more challenging for the LLM to solve. In addition, we observe that the Z3 execution rate on the FOLIO dataset is relatively low, which partially explains why the performance advantage of SSR over CoT is less pronounced on FOLIO.
\begin{table}
\centering
\caption{In the Qwen2.5-7B-Instruct model, we counted the number of right/wrong (r/w) samples processed by the Z3 symbolic solver, as well as the number of right/wrong samples processed by CoT due to excessive logical language translation errors in SSR. These data were then treated as a 2 $\times$ 2 contingency table to calculate the $\chi^2$ values in the independence test respectively.}
\small
\begin{tabular}{l|cc|c}
\toprule
Dataset & Symbol (r/w) & CoT (r/w) & $\chi^2$ \\
\midrule
PrOntoQA & 194/4 & 1/1 & 18.70 \\
Proofwriter & 149/9 & 23/19 & 43.09 \\
FOLIO & 74/19 & 63/43 & 9.36 \\
LogicalDeduction & 124/28 & 26/22 & 14.62 \\
\bottomrule
\end{tabular}
\vspace{-0.3cm}
\label{tab: chi}
\end{table}

\begin{figure}
    \centering
    \includegraphics[width=1.0\linewidth]{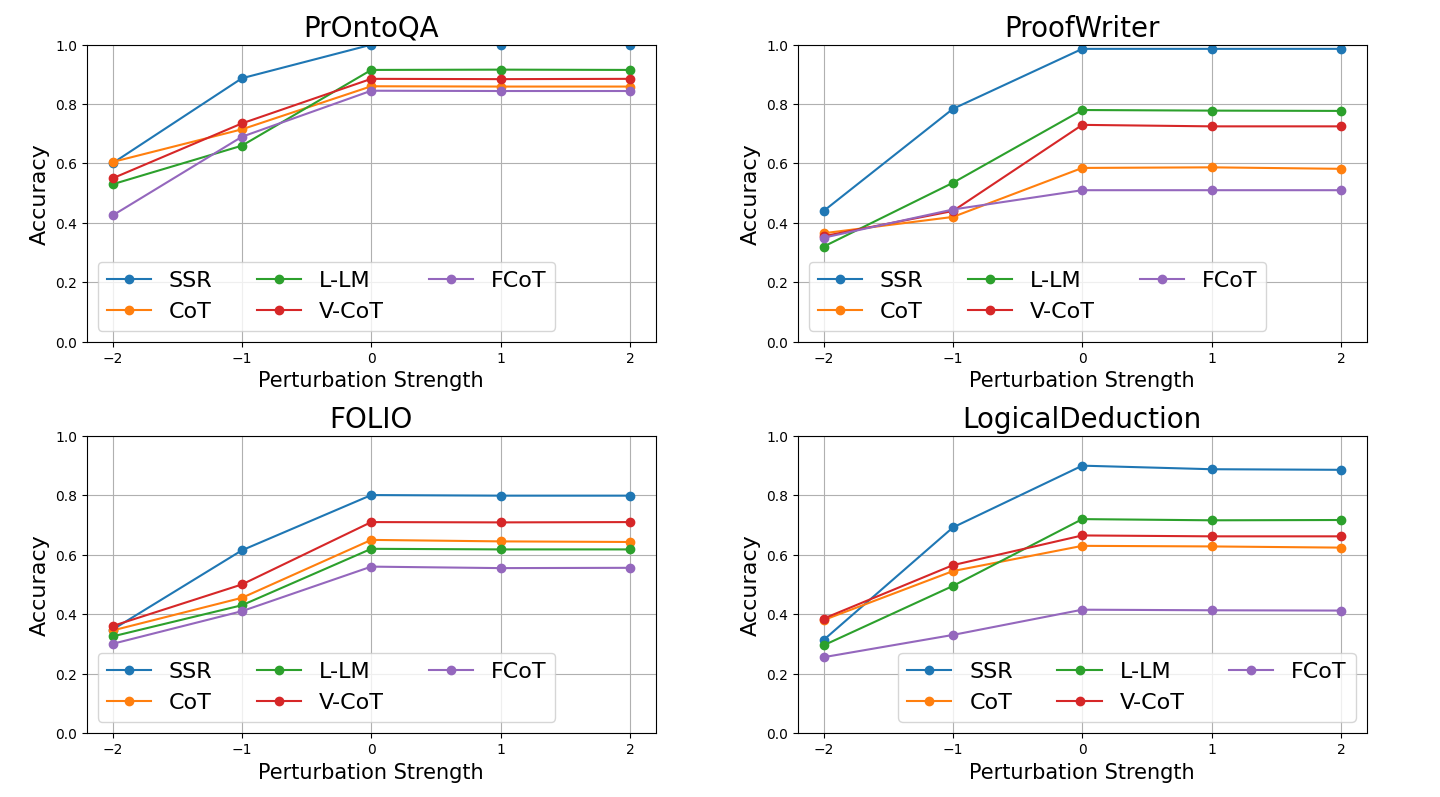}
    \caption{Comparative study on different datasets using the Qwen-2.5-7B-Instruct model. Where the horizontal axis (perturbation strength) denotes the number of randomly added perturbing words or phrase, and the vertical axis denotes reasoning accuracy.}
    \label{fig:robust}
    \vspace{-0.4cm}
\end{figure}

\noindent \textbf{Robustness under Input Perturbations.} In addition, we evaluate the robustness of SSR under controlled perturbations. Specifically, the setting $0$ denotes the original input, while $+1$ and $+2$ indicate the injection of one or two distractor facts, respectively. Conversely, $-1$ and $-2$ correspond to the removal or modification of one or two key premises. For each selected premise, we randomly choose with equal probability whether to delete it or modify it; in the latter case, its value is replaced by another candidate function value sampled uniformly at random, e.g., changing $\mathrm{Pos}[\text{``Cat''}]=2$ to $\mathrm{Pos}[\text{``Cat''}]=3$. Figure~\ref{fig:robust} shows that, under the removal/modification settings, all methods experience performance degradation. Nevertheless, SSR consistently exhibits higher accuracy and a smoother degradation trend than baseline methods, indicating stronger robustness to both distractor information and incomplete or corrupted premises. This stability stems from the integration of symbolic constraints with soft logic, which helps preserve correct logical relations, dependencies, and sufficiency under noisy or partially perturbed inputs. We also observe that adding irrelevant conditions has a relatively minor impact, whereas removing or modifying given conditions leads to a substantially larger performance drop. The soft-logic protection in SSR performs better than the baselines under the $-1$ setting, but this advantage becomes less pronounced under the $-2$ setting.

\begin{table}
\vspace{-0.4cm}
\centering
\small
\caption{Ablation experiments on partial components of SSR with the Qwen2.5-7B-Instruct model. Where ``w/o Filter'' refers to the variant of SSR without the module for filtering obviously erroneous translated logical expressions, ``w/o ECW'' denotes the variant where the soft logic module of SSR assigns identical rather than entropy-based confidence weights to all conditions.}
\begin{tabular}{l|c|c|c}
\toprule
Dataset & w/o Filter & Standard & w/o ECW \\
\midrule
PrOntoQA & 97.0 & \textbf{97.5} & \textbf{97.5}\\
Proofwriter & 81.5 & \textbf{86.0} & 84.5\\
FOLIO & 48.0 & \textbf{64.5} & 64.0\\
LogicalDeduction & 68.5 & \textbf{75.0} & \textbf{75.0}\\
\bottomrule
\end{tabular}
\label{tab:ablation}
\vspace{-0.6cm}
\end{table}

\noindent \textbf{Ablation Study on Key Components.} 
Table~\ref{tab:ablation} presents an ablation study that highlights the contributions of key components in the SSR framework. Removing the filtering module leads to consistent performance degradation across all datasets, indicating that explicitly eliminating obviously erroneous logical expressions translated by LLMs is crucial for maintaining reliable reasoning accuracy. This effect is particularly evident on datasets requiring more complex multi-step deductions. In contrast, incorporating entropy-based confidence weighting in the soft logic module yields relatively modest improvements, suggesting that while uniform weighting already provides a strong baseline, adaptive weighting further refines the aggregation of logical conditions (It may mainly take effect in edge cases where the uniform confidence weights corresponding to two candidate outcomes are the same). Overall, these results demonstrate that both the filtering mechanism and the soft logic design play complementary roles.
\section{Related Work}

\paragraph{Logical Reasoning of LLMs and soft logic.}
LLMs have demonstrated strong empirical performance across a wide range of tasks, yet logical reasoning remains particularly challenging. Logical reasoning requires maintaining globally consistent symbolic states, correctly handling logical relations, dependencies, and sufficiency~\cite{plaat2024reasoning}. To address these challenges, prior work has explored several directions, including prompt-based methods such as chain-of-thought (CoT)~\cite{wei2022chain}, self-consistency~\cite{wang2022self}, neuro-symbolic systems~\cite{pan2023logic,feng2025vericot}, supervised fine-tuning~\cite{luong2024reft,yu2025long}, and implicit reasoning approaches that internalize reasoning processes without explicit intermediate steps~\cite{hao2024training,shen2025codi}.

Despite these advances, LLM-based logical reasoning remains fragile in the presence of noisy inputs, or intermediate reasoning errors, raising significant robustness concerns~\cite{zhou2024can}. Existing robustness-oriented methods either focus on error localization and self-feedback, where models iteratively revise incorrect premises or reasoning steps based on verification signals~\cite{sun2025improving}, or adopt soft logical relaxation~\cite{dubois1998soft,bach2017hinge,diligenti2017semantic}, which allows symbolic reasoning rules to be satisfied in a graded rather than strictly binary manner, improving robustness when the input facts or intermediate predictions are uncertain.

\paragraph{Neuro-Symbolic Reasoning and Auto-Formalization.}
Neuro-symbolic systems seek to combine the flexibility of neural language understanding with the rigor of symbolic reasoning~\cite{hitzler2022neuro,hitzler2022neuro2}. A common pipeline involves translating natural-language premises into formal logic (\emph{auto-formalization}), followed by symbolic inference using a solver. Representative approaches include directly solving translated logic programs~\cite{pan2023logic}, verifying LLM-generated reasoning chains step by step~\cite{feng2025vericot}, or perform knowledge graph reasoning via a differentiable model~\cite{shengyuan2023differentiable}. Despite their promise, these systems face a critical bottleneck in auto-formalization: The neural system can hardly avoid errors and incompleteness in the translation of natural language, whereas symbolic logic requires precise and complete formal representations. Minor errors in translation can easily render the symbolic problem unsatisfiable or misleading. Classical neuro-symbolic reasoning frameworks therefore either assume near-perfect translations or rely on repeated LLM-solver interaction to correct contradictions, which can be brittle or inefficient.


\section{Conclusion}

In this paper, we introduced Symbolic-Neural Soft-Logic Reasoning (SSR), a unified neuro-symbolic framework designed to address the fragility and readability limitations of neuro-symbolic systems. By integrating soft logic constraints, SSR enables verifiable yet flexible reasoning under uncertainty/perturbations and fosters deeper interaction between symbolic and neural components. This design not only improves reasoning accuracy, but also supports the automatic generation of human-like and logically verifiable reasoning chains for training and fine-tuning. Through extensive experiments, we demonstrate that SSR consistently outperforms existing neuro-symbolic and CoT-based approaches while maintaining a certain degree of robustness to added, removed, or modified perturbations during reasoning. Overall, we believe SSR provides a reliable and scalable foundation for advancing trustworthy reasoning and automatically obtain verifiable reasoning-chain data.

\clearpage
\section*{Limitations}
Despite the effectiveness of SSR, our method still has several limitations. First, LLMs may still introduce new errors when translating natural language into formal logical representations. Although the proposed soft-logic mechanism alleviates this issue to some extent, it does not fundamentally eliminate translation errors. We therefore expect future work to further improve the reliability and accuracy of neural systems in generating formal logical representations.

Second, SSR uses a filtering mechanism to remove obviously incorrect formal logical statements, and the dual-system interaction can help address missing conditions or commonsense information. However, the current targeted retrieval threshold is limited to a single conditional statement. When this threshold is exceeded, SSR falls back to CoT reasoning. This safeguard prevents symbolic reasoning from collapsing under excessive noise, but it also introduces a dependency on conventional reasoning paths and fails to exploit the other formal logical statements that have already been obtained.

Third, although SSR can provide verifiable and automatically generated chain-of-thought data for training LLMs, it still shares a common limitation of neuro-symbolic systems: neural models may bypass key reasoning steps instead of internalizing the intended symbolic reasoning process.

\section*{Ethical Considerations}
This paper aims to improve the reasoning performance and robustness of large language models by integrating neuro-symbolic modeling and soft logic, and to empower further development of deep neural networks through structured symbolic systems. From a societal perspective, our work is primarily applicable to areas such as trustworthy automated reasoning, AI planning, and AI for mathematics and science. We do not anticipate negative ethical or societal impacts arising from these applications, as the proposed methods are designed to enhance reliability and interpretability rather than to introduce new risks.


\bibliography{references}

\newpage
\cleardoublepage
\thispagestyle{empty}
\newpage
\appendix

\section{Details of Logical Formalization}
\label{app: symbol}

This appendix provides the technical details omitted from the main paper, including the formalization rules used for translating natural language into symbolic logic and the exact reasoning procedures employed by the SMT solver.

\subsection{Object and Predicate Abstraction}

Given a natural language statement, we instruct the LLM to identify all salient \textbf{objects}, defined as entities that possess attributes or participate in relations. Properties of objects are represented using a unary predicate abstraction $\text{Predicate}(\text{object})$.

Predicates are categorized according to their return types:
\begin{itemize}
    \item \textbf{Numeric Sort} (\texttt{RealSort} and \texttt{IntSort}), which return numerical values
    and support arithmetic constraints (e.g., $\text{Age}(\text{Bob}) = 16$). The domain of these predicate variables is assumed to be a finite, enumerable subset of the real numbers, which in practice is usually instantiated as integers.
    \item \textbf{Boolean Sort} (\texttt{BoolSort}), which return truth values and represent states or conditions
    (e.g., $\text{Happy}(\text{Anne})$). Boolean predicates are assumed to evaluate to \texttt{True}
    unless explicitly negated. Predicate variables of type \texttt{BoolSort} are restricted to the domain $\{\texttt{True}, \texttt{False}\}$. Boolean predicates are assumed to evaluate to \texttt{True} unless explicitly negated.
\end{itemize}

All predicates are normalized to a consistent tense and grammatical number. For relations involving multiple objects, only the primary object appears as the predicate argument, while auxiliary objects are encoded in the predicate name. For example, ``The cat chases the mouse'' is translated as $\text{Chase\_mouse}(\text{cat})$,
and ``Every classmate likes Jane'' is translated as 
\[
\forall x.\ \text{Classmate\_of\_Jane}(x) \rightarrow \text{Like\_Jane}(x).
\]

Compared to multi-arity predicate representations, this design reduces errors caused by incorrect argument ordering or formatting inconsistencies in LLM outputs, at the cost of longer predicate names and reduced expressiveness.

\subsection{Logic Operators and Precedence}

The LLM is restricted to a predefined set of symbolic logical operators, including $\lnot$, $\land$, $\lor$, $\rightarrow$, $\leftrightarrow$, $\forall$, and $\exists$.

Operator precedence follows standard first-order logic conventions, with parentheses used to override default evaluation order. The complete operator precedence table and syntax constraints are enforced during symbolic parsing to ensure compatibility with SMT solvers.
\begin{enumerate}
    \item Numeric Predicates ($\text{Age}(\text{Bob})$)
    \item Arithmetic Operators ($**,*, //,/, +, -$, arranged in accordance with operator precedence.)
    \item Relational Operators ($<, >, \leq, \geq, =, \neq$)
    \item Boolean Predicates ($\text{Happy}(\text{Anne})$)
    \item Logical Negation ($\neg$)
    \item Logical Conjunction ($\land$)
    \item Logical Disjunction ($\lor$)
    \item Implication ($\rightarrow$)
    \item Quantifiers ($\forall, \exists$)
\end{enumerate}

\subsection{Output Structure}

The LLM outputs a structured JSON object with three keys:
\texttt{objects}, \texttt{facts}, and \texttt{query},
each mapping to a list of symbolic expressions represented as strings.
This structured output is directly consumed by the symbolic solver.

\section{SMT-based Reasoning Procedures}
\label{app: procedure}

We now formally describe how the symbolic facts and queries are processed by an SMT solver.

\subsection{Satisfiability Modulo Theories}

Let the set of logical facts be
\begin{equation}
\Phi = \varphi_1 \wedge \varphi_2 \wedge \cdots \wedge \varphi_n,
\end{equation}
and let the query set be $Q = \{q_1, q_2, \ldots, q_m\}$.

Let the predicate variables appearing in $\Phi$ and $Q$ be denoted by $[a_1, a_2, \ldots, a_k]$, where each $a_j$ ranges over a domain $\mathcal{D}_j$. The Cartesian product
\begin{equation}
D = \mathcal{D}_1 \times \mathcal{D}_2 \times \cdots \times \mathcal{D}_k
\end{equation}
represents the space of admissible assignments.

A finite set of formulas $\{p_1, \ldots, p_w\}$ is said to be \emph{satisfiable}
(\textbf{sat}) if there exists an assignment $\alpha \in D$ such that
\begin{equation}
p_1(\alpha) \wedge \cdots \wedge p_w(\alpha) = \texttt{True}.
\end{equation}
Otherwise, it is \emph{unsatisfiable} (\textbf{unsat}).

\subsection{Boolean Query Entailment}

For a closed Boolean query $q_i$, entailment is decided via two satisfiability checks:
\[
\Phi \wedge q_i
\quad \text{and} \quad
\Phi \wedge \lnot q_i.
\]

The result is interpreted as follows:
\begin{itemize}
    \item If $\Phi \wedge q_i$ is \textbf{sat} and $\Phi \wedge \lnot q_i$ is \textbf{unsat},
    then $\Phi \models q_i$.
    \item If $\Phi \wedge q_i$ is \textbf{unsat} and $\Phi \wedge \lnot q_i$ is \textbf{sat},
    then $\Phi \models \lnot q_i$.
    \item If both are \textbf{sat}, the truth of $q_i$ is undetermined under $\Phi$.
\end{itemize}

Note that $\Phi$ itself must be satisfiable for any well-posed reasoning task. By the law of excluded middle, it is impossible for both $\Phi \wedge q_i$ and $\Phi \wedge \lnot q_i$ to be unsatisfiable simultaneously.

\subsection{Solving for Numeric Predicate Variables}

If a query contains a numeric predicate variable $a_i$ to be solved, we enumerate candidate values $a_i \in \{a_{i1}, a_{i2}, \ldots, a_{i t_i}\}$. For each candidate value $a_{ij}$, we check whether $\Phi \wedge (a_i = a_{ij})$ is satisfiable.

The solution set is defined as
\begin{equation}
\mathrm{Sol}(a_i) = \{\, v \in \mathcal{D}_i \mid \Phi \wedge (a_i = v)\ \text{is \textbf{sat}} \,\}.
\end{equation}

\subsection{Arithmetic Computation and Verification}

When a problem consists solely of algebraic computations rather than formal logical reasoning, the framework effectively reduces to a specialized form of \emph{Program-of-Thoughts}. In such cases, the LLM translates natural language descriptions into algebraic constraints, and the symbolic solver deterministically computes the solution. For example, consider the fact set
\begin{equation}
\text{Age(David)} = \text{Age(Peter)} + 3,
\end{equation}
\begin{equation}
\text{Age(David)} - 10 = 2 \cdot (\text{Age(Peter)} - 10).
\end{equation}
An SMT solver can compute the unique satisfying assignment. Additional commonsense constraints such as $\text{Age(David)} > 0$ and
$\text{Age(Peter)} > 0$ can be incorporated to verify plausibility.

This satisfiability-based formulation also supports tasks such as automated theorem proving, counterexample construction, and feasible instance generation in combinatorial reasoning problems.

\section{SSR Soft-Logic Reasoning Algorithm}
Below, we present the pseudocode for the SSR algorithm (Algorithm~\ref{alg:ssr}).

\begin{algorithm}[t]
\caption{Symbolic-Neural Soft-Logic Reasoning (SSR)}
\label{alg:ssr}
\begin{algorithmic}[1]

\STATE \textbf{Input:} Fact set $\Phi = \{\varphi_1,\dots,\varphi_n\}$ translated from LLM, query $q$
\STATE \textbf{Output:} Answer to query $q$

\STATE \textbf{// Syntax-level filtering}
\STATE For each $\varphi \in \Phi$:
\STATE \quad If $\varphi$ is syntactically malformed, replace it with $\mathtt{BoolVal(True)}$.

\STATE If the number of $\mathtt{BoolVal(True)} > 1$, or $q = \mathtt{BoolVal(True)}$:
\STATE \quad \textbf{return} fallback to CoT-based reasoning.

\STATE \textbf{// Confidence-weighted softening}
\STATE For each $\varphi \in \Phi$, assign a confidence weight $w(\varphi) > 0$.

\STATE \textbf{// Enumerate candidate query values}
\STATE Let $\mathcal{Q} = \{q_1, q_2, \dots\}$ be the candidate values for $q$.
\STATE For each $q_j \in \mathcal{Q}$:
\STATE \quad Check satisfiability of $\Phi \wedge (q = q_j)$
\STATE \quad Record the result as $\mathrm{sat}$ or $\mathrm{unsat}$.

\STATE \textbf{// Case I: exactly one satisfiable result}
\STATE If exactly one $q_j$ yields $\mathrm{sat}$:
\STATE \quad \textbf{return} $q = q_j$.

\STATE \textbf{// Case II: no satisfiable result}
\STATE If no $q_j$ yields $\mathrm{sat}$:
\STATE \quad For each $q_j \in \mathcal{Q}$, find a subset $\Phi'_j \subseteq \Phi$ maximizing
\STATE \quad $\sum_{\varphi \in \Phi'_j} w(\varphi)$ such that $\Phi'_j \wedge (q = q_j)$ is $\mathrm{sat}$.
\STATE \quad Select the $q_j$ with the maximum total weight.
\STATE \quad If multiple candidates tie, \textbf{return} \texttt{unknown};
\STATE \quad otherwise, \textbf{return} $q = q_j$.

\STATE \textbf{// Case III: multiple satisfiable results}
\STATE While multiple $q_j$ yield $\mathrm{sat}$:
\STATE \quad Enumerate candidate predicate values.
\STATE \quad For each candidate value:
\STATE \qquad If fixing $a = \text{value}$ yields exactly one $\mathrm{sat}$ result,
\STATE \qquad replace it by a stronger derivable assignment if possible and continue the iteration.
\STATE \quad Stop when no further refinement is possible.

\STATE Ask the LLM to verify whether the derived predicate assignments $a = \text{value}$ follow from the problem statement or commonsense.
\STATE If the LLM confirms:
\STATE \quad \textbf{return} the unique solution;
\STATE Otherwise:
\STATE \quad \textbf{return} \texttt{unknown}.

\end{algorithmic}
\end{algorithm}

\section{Additional Experimental Details}
\label{app: experiment}

\subsection{Prompts Used in Experiments}

This subsection reports the exact prompts used for different reasoning methods in our experiments to ensure reproducibility and fair comparison.

\paragraph{Direct Answer Prompt.}
For the direct reasoning baseline, we instruct the model to solve the problem and output only the final answer without intermediate reasoning:
\begin{Verbatim}[breaklines, breakanywhere]
You are a logical reasoning expert.
Solve the multiple-choice question and directly output the final answer format exactly as: '#### <Letter>'.
\end{Verbatim}

\paragraph{Chain-of-Thought (CoT) Prompt.}
For the CoT baseline, we encourage step-by-step reasoning while enforcing a strict final answer format:
\begin{Verbatim}[breaklines, breakanywhere]
You are a logical reasoning expert. 
Solve the multiple-choice question step by step and explain your reasoning in a simple structured way.
Avoid overly long explanations. Provide clear reasoning followed by the final answer.
At the very end of your response, output the final answer format exactly as: '#### <Letter>'. 
Example: '#### A' or '#### B'.
\end{Verbatim}

\paragraph{SSR Prompt for Boolean logical deduction Tasks.}
For SSR on Boolean logical deduction datasets, we prompt the model to translate the problem into a formal logical representation in JSON format:
\begin{Verbatim}[breaklines, breakanywhere]
You are a formal-logic translator.
Return ONLY a valid JSON object with exactly these keys: "objects", "facts", "query".
FORMALIZATION RULES:
1. "objects": All names of whose properties are described in the context. Note that group-referrig nouns like "Zumpus" and "Vumpus" are predicates rather than objects. That is to say, words like "Dumpus" and "Rompus" cannot be used after predicates and cannot be added into "objects".
2. "facts": Each element must be a two-element list: ["<ORIGINAL_SENTENCE>", "<LOGICAL_FORM>"]. <LOGICAL_FORM> must be:
- An atomic predicate with exactly one object inside parentheses. All objects must be in the singular form.
- Use a prefix "not " for negation if and only if the natural language explicitly indicates negation. e.g. "not Happy(Anne)".
- Use "->" to represent the logical relationship of "if...then...", for example: ["If Anne is green then Anne is round.", "Green(Anne) -> Round(Anne)"]
- Use "and", "or" for a boolean combination of atomic predicates. e.g. "Have_a_cat(Bob) and Tall(Bob)", "not (Happy(Anne) or Sad(Anne))".
- For universal quantified laws where the object is not specified, use the form: "forall x. <premise> -> <conclusion>". Example: ["Tumpuses are wumpuses", "forall x. Tumpus(x) -> Wumpus(x)"], ["Every tumpus is large", "forall x. Tumpus(x) -> Large(x)"]
3. "query": a single string representing the formula to evaluate. Do NOT include the query in the "facts" list.
4. Permitted logical terms (only): and, or, ->, forall, exist, not, and Predicate(Object)
Translate the following context and question into logical language. For the "facts" section, the "original sentence" must be copied verbatim sentence by sentence in the exact order of the original context without any modifications; the "logical formalization" must be translated rigorously in accordance with the original sentences. Do not omit, add, or alter any conditions.
Context:
{context}
Question:
{question}
\end{Verbatim}

\paragraph{SSR Prompt for Numeric Logical Reasoning Tasks (LogicalDeduction).}
For numeric-based reasoning tasks like ordering and integer programming, we use a specialized formalization prompt with position-based predicates:
\begin{Verbatim}[breaklines, breakanywhere]
You are a formal logic translator.
Convert the following ordering puzzle into a JSON structure.
Return ONLY a valid JSON object with exactly these keys: "objects", "larger_direction", "facts", "query".
Requirements:
1. "objects": Copy the names of all the items to be compared directly as "objects". If the name of an object consists of multiple words, connect them with underscores.
2. "facts": Each element must be a two-element list: ["<ORIGINAL_SENTENCE>", "<LOGICAL_FORM>"]. <LOGICAL_FORM> must use only these predicates and syntax: >, <, =, and, or, not, Pos(object). Comparisons: Pos(A) < Pos(B), Pos(A) > Pos(B), Pos(A) = k. Any other symbols and predicates are prohibited.
3. "query": A dictionary in the form: {{"A": "<logical form of option A>", "B": "<logical form of option B>", "C": "<logical form of option C>", "D": "<logical form of option D>", "E": "<logical form of option E>"}}. Translate each option in the question into the corresponding logical language.
Note:
You must perform sorting comparisons in the following order: left < right, bottom < top, front < back and small < large (e.g., new < old, cheap < expensive).  For example, if there are 5 objects arranged from left to right and "apple is the second from the right", you should translate it as "Pos(apple) = 4" based on the total number of objects. All your comparison criteria must be consistent; additionally, pay attention to the impact of words like "less" and "more" on the translation into logical language.
Context
{context}
Question
{question}
Begin.
\end{Verbatim}

\subsection{Perturbation Sentences Used in Robustness Experiments}

To evaluate robustness under irrelevant or noisy inputs, we augment the original problem context with a fixed set of neutral perturbation sentences. These sentences are intentionally designed to be semantically self-contained, logically valid, and largely unrelated to the target reasoning task. They include general factual statements, commonsense knowledge, meta-level comments, numerical-information-based perturbations, and simple tautologies. Importantly, none of these perturbations introduces new task-relevant constraints or logical dependencies. This design ensures that any performance degradation arises from the sensitivity of a model to distracting information, rather than changes to the underlying logical structure of the problem. If not handled properly, these newly added sentences can introduce additional errors into SSR, as their corresponding logical translations may contain formatting errors, interfere with the LLM’s identification of objects or other facts, or disrupt the retrieval process during automatic reasoning chain generation.

The 20 perturbation sentences used in our experiments are listed below:

\begin{itemize}
    \item This sentence is included for completeness but does not affect the problem.
    \item The following text contains multiple independent statements.
    \item Logical reasoning tasks may vary in difficulty.
    \item Please note that not all sentences are equally important.
    \item Water boils at 100 degrees Celsius at sea level.
    \item The Earth completes one full rotation every 24 hours.
    \item A standard chessboard has 64 squares.
    \item Most modern buildings are made of concrete or steel.
    \item It was raining lightly outside the room.
    \item The room was quiet except for the sound of a ticking clock.
    \item After a long day, the reader might feel slightly tired.
    \item The experiment was conducted late in the afternoon.
    \item The meeting lasted for 45 minutes.
    \item There are seven continents on Earth.
    \item The building has three elevators and twelve floors.
    \item A triangle has exactly three sides.
    \item If today is Monday, then tomorrow is Tuesday.
    \item All squares are rectangles, but not all rectangles are squares.
    \item If A implies B, the reverse does not necessarily hold.
    \item Either the statement is true, or it is false.
\end{itemize}

\section{Case Studies and Further Discussions}
\label{app: case}
Below, we provide several specific examples from the main experiment (Table~\ref{tab: accuracy}) and a deeper analysis of the effectiveness of our SSR method.

\subsection{PrOntoQA Dataset on Qwen-2.5-7B}
\begin{tcolorbox}

\textbf{Question: }``Jompuses are large. Every jompus is a zumpus. Each zumpus is sweet. Zumpuses are numpuses. Every numpus is hot. Each tumpus is opaque. Numpuses are yumpuses. Most modern buildings are made of concrete or steel. Every yumpus is brown. Each yumpus is a wumpus. Wumpuses are not opaque. Wumpuses are impuses. Fae is a jompus. Is the following statement true or false? Fae is opaque."

\textbf{Correct Answer: }B (False)

\textbf{LLM's Answer: }B (False)

\textbf{Objects: }["Fae"]

\textbf{Facts: }["forall x. Jompus(x) $\rightarrow$ Large(x)", "forall x. Jompus(x) $\rightarrow$ Zumpus(x)", "forall x. Zumpus(x) $\rightarrow$ Sweet(x)", "forall x. Zumpus(x) $\rightarrow$ Numpus(x)", "forall x. Numpus(x) $\rightarrow$ Hot(x)", "forall x. Tumpus(x) $\rightarrow$ Opaque(x)", "forall x. Numpus(x) $\rightarrow$ Yumpus(x)", "forall x. (Modern(x) and Building(x)) $\rightarrow$ (Made\_of\_concrete(x) or Made\_of\_steel(x))", "forall x. Yumpus(x) $\rightarrow$ Brown(x)", "forall x. Yumpus(x) $\rightarrow$ Wumpus(x)", "forall x. Wumpus(x) $\rightarrow$ not Opaque(x)", "forall x. Wumpus(x) $\rightarrow$ Impus(x)", "Jompus(Fae)"]

\textbf{Query: }["Opaque(Fae)"]

\textbf{Sat\_query: }False

\textbf{Sat\_not\_query: }True

\textbf{Forward: }["Jumpus(Fae) $\rightarrow$ Zumpus(Fae)", "Zumpus(Fae) $\rightarrow$ Numpus(Fae)", "Numpus(Fae) $\rightarrow$ Yumpus(Fae)", "Yumpus(Fae) $\rightarrow$ Wumpus(Fae)", "Wumpus(Fae) $\rightarrow$ not Opaque(Fae)"]

\textbf{Correct\_forward: }True

\textbf{Backward: }["not Opaque(Fae) $\leftarrow$ Wumpus(Fae)", "Wumpus(Fae) $\leftarrow$ Yumpus(Fae)", "Yumpus(Fae) $\leftarrow$ Numpus(Fae)", "Numpus(Fae) $\leftarrow$ Zumpus(Fae)", "Zumpus(Fae) $\leftarrow$ Jumpus(Fae)"]

\textbf{Correct\_backward: }True
\end{tcolorbox}

\begin{tcolorbox}

\textbf{Question: }``Rompuses are spicy. Every rompus is an impus. Yumpuses are not small. Impuses are orange. Impuses are zumpuses. Zumpuses are not hot. Zumpuses are numpuses. Numpuses are metallic. Numpuses are wumpuses. Every wumpus is not kind. Each wumpus is a dumpus. Each dumpus is not bright. Every dumpus is a jompus. Jompuses are small. Jompuses are vumpuses. Each vumpus is not shy. Every vumpus is a tumpus. It was raining lightly outside the room. Alex is a zumpus. Is the following statement true or false? Alex is not small."

\textbf{Correct Answer: }B (False)

\textbf{LLM's Answer: }C (Unknown)

\textbf{Objects: }["Alex", "Rompus", "Impus", "Yumpus", "Zumpus", "Numpus", "Wumpus", "Dumpus", "Jompus", "Vumpus", "Tumpus"]

\textbf{Facts: }["Spicy(Rompus)", "forall x. Rompus(x) $\rightarrow$ Impus(x)", "not Small(Yumpus)", "Orange(Impus)", "forall x. Impus(x) $\rightarrow$ Zumpus(x)", "forall x. Zumpus(x) $\rightarrow$ not Hot(Zumpus)", "forall x. Zumpus(x) $\rightarrow$ Numpus(x)", "forall x. Numpus(x) $\rightarrow$ Metallic(Numpus)", "forall x. Numpus(x) $\rightarrow$ Wumpus(x)", "forall x. Wumpus(x) $\rightarrow$ not Kind(Wumpus)", "forall x. Wumpus(x) $\rightarrow$ Dumpus(x)", "forall x. Dumpus(x) $\rightarrow$ not Bright(Dumpus)", "forall x. Dumpus(x) $\rightarrow$ Jompus(x)", "forall x. Jompus(x) $\rightarrow$ Small(Jompus)", "forall x. Jompus(x) $\rightarrow$ Vumpus(x)", "forall x. Vumpus(x) $\rightarrow$ not Shy(Vumpus)", "forall x. Vumpus(x) $\rightarrow$ Tumpus(x)", "Raining\_lightly\_outside\_the\_room()", "Zumpus(Alex)"]

\textbf{Query: }["not Small(Alex)"]

\textbf{Sat\_query: }True

\textbf{Sat\_not\_query: }True

\textbf{Forward: }[]

\textbf{Correct\_forward: }False

\textbf{Backward: }[]

\textbf{Correct\_backward: }False
\end{tcolorbox}

\subsection{ProofWriter Dataset on Qwen-2.5-7B}

\begin{tcolorbox}

\textbf{Question: }``The bear eats the tiger. The bear is not red. The bear likes the squirrel. Logical reasoning tasks may vary in difficulty. The bear visits the squirrel. The mouse eats the tiger. The mouse is red. The mouse is rough. The mouse likes the bear. The mouse does not like the tiger. The squirrel likes the bear. The tiger eats the mouse. If the mouse likes the bear and the bear likes the mouse then the mouse is not kind. If someone visits the mouse then they like the squirrel. If someone is green then they visit the bear. If someone likes the bear and the bear visits the tiger then they visit the bear. If someone eats the bear and they do not like the tiger then they are green. If someone visits the mouse then the mouse eats the bear. If someone is red and they eat the tiger then the tiger visits the mouse. If someone visits the bear then they are kind. Based on the above information, is the following statement true, false, or unknown? The tiger does not like the squirrel."

\textbf{Correct Answer: }B (False)

\textbf{LLM's Answer: }C (Unknown)

\textbf{Objects: }["bear", "tiger", "squirrel", "mouse"]

\textbf{Facts: }["Eats(bear, tiger)", "not Red(bear)", "Likes(bear, squirrel)", "Logical\_reasoning\_tasks\_may\_vary\_in\_difficulty.", "Visits(bear, squirrel)", "Eats(mouse, tiger)", "Red(mouse)", "Rough(mouse)", "Likes(mouse, bear)", "not Likes(mouse, tiger)", "Likes(squirrel, bear)", "Eats(tiger, mouse)", "forall x. (Likes(mouse, bear) and Likes(bear, mouse)) $\rightarrow$ not Kind(mouse)", "forall x. Visits(x, mouse) $\rightarrow$ Likes(x, squirrel)", "forall x. Green(x) $\rightarrow$ Visits(x, bear)", "forall x. (Likes(x, bear) and Visits(bear, tiger)) $\rightarrow$ Visits(x, bear)", "forall x. (Eats(x, bear) and not Likes(x, tiger)) $\rightarrow$ Green(x)", "forall x. Visits(x, mouse) $\rightarrow$ Eats(mouse, bear)", "forall x. (Red(x) and Eats(x, tiger)) $\rightarrow$ Visits(tiger, mouse)", "forall x. Visits(x, bear) $\rightarrow$ Kind(x)"]

\textbf{Query: }["not Likes(tiger, squirrel)"]

(Continued on the next page)
\end{tcolorbox}

\begin{tcolorbox}
\small
\textbf{LLM\_CoT: } ``Let's break down the information and analyze the statement "The tiger does not like the squirrel" step by step. 1. **Direct Information:** - The bear eats the tiger. - The bear is not red. - The bear likes the squirrel. - The bear visits the squirrel. - The mouse eats the tiger. - The mouse is red. - The mouse is rough. - The mouse likes the bear. - The mouse does not like the tiger. - The squirrel likes the bear. - The tiger eats the mouse. 2. **Logical Deductions:** - From "The mouse likes the bear" and "The bear likes the squirrel," we can infer that the mouse likes the squirrel (since the bear likes the squirrel and the mouse likes the bear). - From "If the mouse likes the bear and the bear likes the squirrel then the mouse is not kind," we know that the mouse is not kind because it likes the bear and the bear likes the squirrel. - From "If someone visits the mouse then they like the squirrel," we know that the bear likes the squirrel because the bear visits the mouse. - From "If someone is green then they visit the bear," we know that being green is associated with visiting the bear. - From "If someone likes the bear and the bear visits the tiger then they visit the bear," we know that liking the bear and the bear visiting the tiger implies visiting the bear. - From "If someone eats the bear and they do not like the tiger then they are green," we know that eating the bear and not liking the tiger implies being green. - From "If someone visits the mouse then the mouse eats the bear," we know that visiting the mouse implies the mouse eats the bear. - From "If someone is red and they eat the tiger then the tiger visits the mouse," we know that being red and eating the tiger implies the tiger visits the mouse. - From "If someone visits the bear then they are kind," we know that visiting the bear implies being kind. 3. **Analyzing the Statement:** - We need to determine if the tiger does not like the squirrel. - We know the bear likes the squirrel. - The bear visits the squirrel. - The mouse likes the bear. - The mouse does not like the tiger. - The squirrel likes the bear. - The tiger eats the mouse. 4. **Connecting the Dots:** - The bear likes the squirrel. - The mouse likes the bear. - The tiger eats the mouse. - There is no direct information about the tiger's feelings towards the squirrel. 5. **Conclusion:** - We do not have any direct or indirect information that tells us whether the tiger likes or dislikes the squirrel. - Therefore, based on the given information, we cannot determine if the tiger does not like the squirrel. \#\#\#\# C''

\textbf{Forward: }[]

\textbf{Correct\_forward: }False

\textbf{Backward: }[]

\textbf{Correct\_backward: }False
\end{tcolorbox}

\begin{tcolorbox}

\textbf{Question: }``Bob is rough. Charlie is nice. Charlie is rough. Charlie is white. Dave is not white. Gary is cold. Gary is quiet. The following text contains multiple independent statements. Rough, white things are quiet. If Charlie is round and Charlie is quiet then Charlie is rough. All young things are round. Nice things are white. Rough, nice things are cold. If something is round and rough then it is nice. If something is rough then it is young. Based on the above information, is the following statement true, false, or unknown? Gary is not white."

\textbf{Correct Answer: }C (Unknown)

\textbf{LLM's Answer: }C (Unknown)

\textbf{Objects: }["Bob", "Charlie", "Dave", "Gary"]

\textbf{Facts: }["Rough(Bob)", "Nice(Charlie)", "Rough(Charlie)", "White(Charlie)", "not White(Dave)", "Cold(Gary)", "Quiet(Gary)", "forall x. (Rough(x) and White(x)) $\rightarrow$ Quiet(x)", "forall x. (Round(x) and Quiet(x)) $\rightarrow$ Rough(x)", "forall x. Young(x) $\rightarrow$ Round(x)", "forall x. Nice(x) $\rightarrow$ White(x)", "forall x. (Rough(x) and Nice(x)) $\rightarrow$ Cold(x)", "forall x. (Round(x) and Rough(x)) $\rightarrow$ Nice(x)", "forall x. Rough(x) $\rightarrow$ Young(x)"]

\textbf{Query: }["not White(Gary)"]

\textbf{Sat\_query: }True

\textbf{Sat\_not\_query: }True

\textbf{Forward: }(Terminate before reaching the query according to the procedure)

\textbf{Correct\_forward: }True

\textbf{Backward: }["White(Gary) $\leftarrow$ Nice(Gary)", "Nice(Gary) $\leftarrow$ (Round(Gary) and Rough(Gary))", "Round(Gary) $\leftarrow$ Young(Gary)", "Young(Gary) $\leftarrow$ Rough(Gary)", "Rough(Gary) $\leftarrow$ (Round(Gary) and Quiet(Gary))", "Round(Gary) $\leftarrow$ Young(Gary)", ...] (infinite loop)

\textbf{Correct\_backward: }False
\end{tcolorbox}

\subsection{FOLIO Dataset on Qwen-2.5-7B}
\begin{tcolorbox}
\small
\textbf{Question: }``"There are six types of wild turkeys: Eastern wild turkey, Osceola wild turkey, Gould’s wild turkey, Merriam’s wild turkey, Rio Grande wild turkey, and Ocellated wild turkey. Most modern buildings are made of concrete or steel. Tom is not an Eastern wild turkey. Tom is not an Osceola wild turkey. Tom is also not a Gould's wild turkey, or a Merriam's wild turkey, or a Rio Grande wild turkey. Tom is a wild turkey. Based on the above information, is the following statement true, false, or uncertain? Tom is an Ocellated wild turkey.""

\textbf{Correct Answer: }A (True)

\textbf{LLM's Answer: }A (True)

\textbf{Objects: }["Tom"]

\textbf{Facts: }["forall x. Wild\_turkey(x) $\rightarrow$ (Eastern\_wild\_turkey(x) or Osceola\_wild\_turkey(x) or Goulds\_wild\_turkey(x) or Merriams\_wild\_turkey(x) or Rio\_Grande\_wild\_turkey(x) or Ocellated\_wild\_turkey(x))", "forall x. (Modern\_buildings(x) $\rightarrow$ (Concrete(x) or Steel(x)))", "not Eastern\_wild\_turkey(Tom)", "not\ Osceola\_wild\_turkey(Tom)", "not Goulds\_wild\_turkey(Tom) and not Merriams\_wild\_turkey(Tom) and not Rio\_Grande\_wild\_turkey(Tom)", "Wild\_turkey(Tom)"]

\textbf{Query: }["Wild\_turkey(Tom) and Ocellated\_wild\_turkey(Tom)"]

\textbf{Sat\_query: }True

\textbf{Sat\_not\_query: }False

\textbf{Forward: }["Wild\_turkey(Tom) $\rightarrow$ (Eastern\_wild\_turkey(Tom) or Osceola\_wild\_turkey(Tom) or Goulds\_wild\_turkey(Tom) or Merriams\_wild\_turkey(Tom) or Rio\_Grande\_wild\_turkey(Tom) or Ocellated\_wild\_turkey(Tom))", ""not Eastern\_wild\_turkey(Tom)", "not\ Osceola\_wild\_turkey(Tom)", "not Goulds\_wild\_turkey(Tom) and not Merriams\_wild\_turkey(Tom) and not Rio\_Grande\_wild\_turkey(Tom)" $\rightarrow$ Ocellated\_wild\_turkey(Tom)"]

\textbf{Correct\_forward: }True

\textbf{Backward: }["Wild\_turkey(Tom) $\leftarrow$ Wild\_turkey(Tom)", "Ocellated\_wild\_turkey(Tom) $\leftarrow$ "forall x. Wild\_turkey(x) $\rightarrow$ (Eastern\_wild\_turkey(x) or Osceola\_wild\_turkey(x) or Goulds\_wild\_turkey(x) or Merriams\_wild\_turkey(x) or Rio\_Grande\_wild\_turkey(x) or Ocellated\_wild\_turkey(x))""] (Terminated)

\textbf{Correct\_backward: }False
\end{tcolorbox}

\begin{tcolorbox}

\textbf{Question: }``"A person is either a Grand Slam champion or an Oscar-nominated actor. All people who are Grand Slam champions are professional tennis players. All Oscar-nominated actors are celebrities. All professional tennis players are athletes. If a person is a celebrity then they are well paid. If a person is an athlete then they are famous. All well-paid people live in tax havens. If Djokovic is famous and is an athlete, then Djokovic lives in well paid. If today is Monday, then tomorrow is Tuesday. Based on the above information, is the following statement true, false, or uncertain? Djokovic lives in a tax haven.""

\textbf{Correct Answer: }A (True)

\textbf{LLM's Answer: }C (Unknown)

\textbf{Objects: }["Djokovic"]

\textbf{Facts: }["forall x. (Grand\_Slam\_champion(x) or Oscaro\_nominated\_actor(x))", "forall x. Grand\_Slam\_champion(x) $\rightarrow$ Professional\_tennis\_player(x)", "forall x. Oscaro\_nominated\_actor(x) $\rightarrow$ Celebrity(x)", "forall x. Professional\_tennis\_player(x) $\rightarrow$ Athlete(x)", "forall x. Celebrity(x) $\rightarrow$ Well\_paid(x)", "forall x. Athlete(x) $\rightarrow$ Famous(x)", "forall x. Well\_paid(x) $\rightarrow$ Lives\_in\_tax\_havens(x)", "forall x. (Famous(x) and Athlete(x)) $\rightarrow$ Lives\_in\_well\_paid(x)", "forall x. Today\_is\_Monday(x) $\rightarrow$ Tomorrow\_is\_Tuesday(x)"]

\textbf{Query: }["Lives\_in\_tax\_havens(Djokovic)"]

\textbf{Sat\_query: }True

\textbf{Sat\_not\_query: }True

\textbf{Discriminator: }["Oscaro\_nominated\_actor\_Djokovic"]

\textbf{Retrieval\_discriminator: }None

\textbf{Forward: }(Terminate before reaching the query according to the procedure)

\textbf{Correct\_forward: }False

\textbf{Backward: }["Lives\_in\_tax\_havens(Djokovic) $\leftarrow$ Well\_paid(Djokovic)", "Well\_paid(Djokovic) $\leftarrow$ Celebrity(Djokovic)", "Celebrity(Djokovic) $\leftarrow$ Oscaro\_nominated\_actor(Djokovic)", "Oscaro\_nominated\_actor(Djokovic) $\leftarrow$ "forall x. (Grand\_Slam\_champion(x) or Oscaro\_nominated\_actor(x))""] (Terminated)

\textbf{Correct\_backward: }False
\end{tcolorbox}

\subsection{LogicalDeduction Dataset on Qwen-2.5-7B}
\begin{tcolorbox}

\textbf{Question: }``Either the statement is true, or it is false. The following paragraphs each describe a set of five objects arranged in a fixed order. The statements are logically consistent within each paragraph. Please note that not all sentences are equally important. On a branch, there are five birds: a cardinal, a robin, a blue jay, a quail, and a raven. The robin is to the right of the raven. The cardinal is the leftmost. It was raining lightly outside the room. The raven is to the right of the blue jay. If A implies B, the reverse does not necessarily hold. The blue jay is the third from the left. A) The cardinal is the second from the right. B) The robin is the second from the right. C) The blue jay is the second from the right. D) The quail is the second from the right. E) The raven is the second from the right."

\textbf{Correct Answer: }E (The raven is the second from the right)

\textbf{LLM's Answer: }E (The raven is the second from the right)

\textbf{Objects: }["cardinal", "robin", "blue\_jay", "quail", "raven"]

\textbf{Facts: }["Pos(robin) $>$ Pos(raven)", "Pos(cardinal) = 1", "Pos(raven) $>$ Pos(blue\_jay)", "Pos(blue\_jay) = 3"]

\textbf{Query: }["Pos(cardinal) = 4", "Pos(robin) = 4", "Pos(blue\_jay) = 4", "Pos(quail) = 4", "Pos(raven) = 4"]

\textbf{Forward: }[""Pos(blue\_jay) = 3", "Pos(robin) $>$ Pos(raven)", "Pos(raven) $>$ Pos(blue\_jay)", $\rightarrow$ Pos(raven) = 4"]

\textbf{Correct\_forward: }True

\textbf{Backward: }["Pos(raven) = 4 $\leftarrow$ "Pos(robin) $>$ Pos(raven)", "Pos(raven) $>$ Pos(blue\_jay)"", "Pos(robin) $>$ Pos(raven) $\leftarrow$ "Pos(robin) = 5"", "Pos(robin) = 5 $\leftarrow$ sat", "Pos(raven) $>$ Pos(blue\_jay) $\leftarrow$ "Pos(blue\_jay) = 3""]

\textbf{Correct\_backward: }True
\end{tcolorbox}

\begin{tcolorbox}

\textbf{Question: }``If today is Monday, then tomorrow is Tuesday. The following paragraphs each describe a set of five objects arranged in a fixed order. Water boils at 100 degrees Celsius at sea level. The statements are logically consistent within each paragraph. Either the statement is true, or it is false. On a branch, there are five birds: a cardinal, a hawk, a hummingbird, a raven, and an owl. The raven is to the left of the hummingbird. The meeting lasted for 45 minutes. The hawk is to the left of the owl. The cardinal is to the left of the hawk. The raven is the second from the right. A) The cardinal is the second from the left. B) The hawk is the second from the left. C) The hummingbird is the second from the left. D) The raven is the second from the left. E) The owl is the second from the left."

\textbf{Correct Answer: }B (The hawk is the second from the left)

\textbf{LLM's Answer: }A (The cardinal is the second from the left)

\textbf{Objects: }["cardinal", "hawk", "hummingbird", "raven", "owl"]

\textbf{Facts: }["Pos(raven) $<$ Pos(hummingbird)", "Pos(hawk) $<$ Pos(owl)", "Pos(cardinal) $<$ Pos(hawk)", "Pos(raven) = 4"]

\textbf{Query: }["Pos(cardinal) = 1", "Pos(hawk) = 1", "Pos(hummingbird) = 1", "Pos(raven) = 1", "Pos(owl) = 1"]

\textbf{Forward: }[""Pos(raven) $<$ Pos(hummingbird)", "Pos(raven) = 4" $\rightarrow$ Pos(hummingbird) = 5", ""Pos(hawk) $<$ Pos(owl)", "Pos(cardinal) $<$ Pos(hawk)", "Pos(raven) = 4" $\rightarrow$ Pos(cardinal) = 1"]

\textbf{Correct\_forward: }False

\textbf{Backward: }[]

\textbf{Correct\_backward: }False
\end{tcolorbox}

\subsection{Failure Analysis of SSR on Enhancing Reasoning Accuracy}

For conciseness and to showcase a diverse set of successful and failed reasoning cases, the examples presented in this section are not randomly selected. We observe that most failures of SSR are closely associated with errors in the LLM-generated formal logical translations. These failures primarily arise from the following causes: (1) incorrect extraction of objects, which subsequently leads to biased interpretation of universally quantified formulas (e.g., the second PrOntoQA example); (2) malformed translations of predicate functions, resulting in invalid or misinterpreted logical expressions (e.g., the first ProofWriter example); (3) improper handling of injected perturbation sentences, which interferes with fact parsing or retrieval (e.g., the first ProofWriter example); (4) lack of alignment between logically equivalent predicate forms expressed with different surface representations (e.g., the second FOLIO example); and (5) incorrect translation of the query itself, causing the downstream reasoning target to be entirely mis-specified (e.g., the second LogicalDeduction example). Although soft-logic relaxation in SSR can mitigate a subset of minor perturbations and translation inconsistencies, the diversity of problem formulations and the inherent stochasticity of LLM outputs make it difficult to fully eliminate all such failure cases.

From a broader perspective, these failure modes can be roughly divided into two categories. Errors caused by surface-level inconsistencies in logical translation, such as malformed predicates, misaligned but semantically equivalent predicate names, or mild interference from perturbation sentences, are in principle repairable through stronger normalization~\cite{zhang2021neural}, constrained decoding~\cite{NEURIPS2024_2bdc2267, geng2023grammar}, or post-hoc correction mechanisms~\cite{feng2025tear}. In contrast, failures stemming from fundamentally incorrect object extraction or mis-specified queries reflect deeper semantic misunderstandings of the problem statement and remain a more systemic challenge. Addressing these issues likely requires improvements in the reliability of LLM-based semantic parsing or tighter coupling between translation and symbolic verification, which we leave for future work.

\subsection{Failure Analysis of Automatic Reasoning Chain Generation}

We further analyze failure cases where SSR fails to automatically generate a coherent human-like reasoning chain, even when it is able to reach a correct or partially correct conclusion. These failures mainly arise from limitations in the retrieval-based forward and backward reasoning procedures. First, incorrect extraction of objects, query or malformed facts can prevent both forward and backward reasoning from performing valid retrieval, as the necessary predicates or relations cannot be matched correctly (e.g., the second PrOntoQA example, the first ProofWriter example, and the second LogicalDeduction example). Second, during backward reasoning, the retrieval process may fail to leverage additional information for elimination, causing the algorithm to enter a dead loop when repeatedly querying the same pair of predicate functions (e.g., the second ProofWriter example). Third, backward reasoning may encounter bottlenecks in problems that require extensive elimination over a large space of candidate values, leading to incomplete or excessively long reasoning chains (e.g., the first FOLIO example). Finally, when the LLM-generated logical translation omits critical conditions, the retrieval-based reasoning process lacks sufficient premises to proceed, resulting in premature termination of the reasoning chain (e.g., the second FOLIO example). These cases highlight that successful chain construction not only depends on correct symbolic inference, but also critically relies on the completeness and structural quality of the underlying logical representation.

\end{document}